\pdfoutput=1
\newcommand*{\ATLASLATEXPATH}{latex/}
\documentclass[UKenglish,texlive=2011,cernpreprint,txfonts]{\ATLASLATEXPATH atlasdoc}

\usepackage{\ATLASLATEXPATH atlaspackage}
\usepackage{placeins}
\usepackage{\ATLASLATEXPATH atlasbiblatex}

\usepackage{\ATLASLATEXPATH atlascontribute}

\usepackage{\ATLASLATEXPATH atlasphysics}

\graphicspath{{logos/}{figures/}}

\usepackage{InclusiveCrossSection2012-defs}

\hypersetup{pdftitle={ATLAS draft},pdfauthor={The ATLAS Collaboration}}

\AtlasTitle{Measurement of the inclusive isolated prompt photon cross
section in $pp$ collisions at $\sqrt{s} = 8~\TeV$ with the
ATLAS detector}

\author{The ATLAS Collaboration}

\AtlasRefCode{STDM-2014-09}

\PreprintIdNumber{CERN-EP-2016-035}

\AtlasJournal{JHEP}
\AtlasJournalRef{JHEP 06 (2016) 005}
\AtlasDOI{10.1007/JHEP08(2016)005}

\AtlasAbstract{%
A measurement of the cross section for the inclusive 
production of isolated prompt photons in proton--proton collisions 
at a centre-of-mass energy of $\sqrt{s} = 8~\TeV$ is presented. 
The measurement covers the pseudorapidity ranges 
$|\etag| < 1.37$ and 
$1.56 \leq |\etag| < 2.37$
in the transverse energy range $25 < \ETg < 1500~\GeV$. 
The results are based on an integrated luminosity of 20.2~\ifb, 
recorded by the ATLAS detector at the LHC.
Photon candidates are identified by combining information from 
the calorimeters and the inner tracker. 
The background is subtracted using a data-driven technique, based on the 
observed calorimeter shower-shape variables and the deposition of hadronic energy in a narrow cone around the photon candidate. The measured cross sections 
are compared with leading-order and next-to-leading order perturbative QCD calculations 
and are found to be in a good agreement over ten orders of magnitude.
}

\AtlasCoverSupportingNote{Supporting Documentation (+files for HepData)}{https://cds.cern.ch/record/1711874}
\AtlasCoverCommentsDeadline{21 February  2016}

\AtlasCoverAnalysisTeam{Brigitte Vachon, Claudia Glasman, Josu Cantero, Juan Terron, Mark Stockton, Michael Stoebe, Othmane Rifki, Sergei Chekanov}

\AtlasCoverEdBoardMember{Bruno~Lenzi, Jonathan~Butterworth~(chair), Leonardo~Carminati, Lidia~Dell'Asta}

\AtlasCoverEgroupEditors{atlas-STDM-2014-09-editors@cern.ch}

\AtlasCoverEgroupEdBoard{atlas-STDM-2014-09-editorial-board@cern.ch}

\begin{document}

\maketitle

\tableofcontents


\section{Introduction}
\label{sec:intro}

Prompt photons, excluding those originating from hadron decays, are produced at the LHC in the hard process 
$pp\rightarrow \gamma +X$. The measurement of this inclusive production provides a probe of perturbative Quantum Chromodynamics 
(pQCD) and specifically, through the dominant leading-order (LO) process $qg\rightarrow q\gamma$, can be used to study the gluon parton distribution function (PDF)~\cite{promptphoton_and_gluon_pdf,Lai:1994bb,Martin:1998sq,Martin:1999ww,PhysRevD.73.094007,dEnterria:2012yj} of the proton. 
In addition, an improved understanding of prompt photon production is potentially important in aiding analyses of processes for which they are an important background (for instance, measurements of the Higgs boson in the diphoton decay channel). 

Inclusive prompt photon production is made up of two contributions: direct and 
fragmentation photons. Direct photons are those associated with the hard sub-process, 
whereas fragmentation photons are produced from the fragmentation of a 
coloured parton. An isolation requirement is used to reduce both the poorly understood 
non-perturbative fragmentation contribution and the contamination from the dominant background of photons originating from hadron decays, mainly light neutral mesons (i.e. $\pi^{0}$, $\eta$).

Inclusive measurements of prompt photons have been made at hadron colliders by ATLAS~\cite{Aad:2010sp,Aad:2011tw,Aad:2013zba}, CMS~\cite{Khachatryan:2010fm,Chatrchyan:2011ue}, CDF~\cite{CDF_promptphoton}, D0~\cite{Abazov:2013pua,D0_promptphoton}, UA1~\cite{Albajar:1988im} and UA2~\cite{Alitti:1991yk}. The analysis presented here uses 20.2~\ifb of proton--proton collision data recorded by the ATLAS detector and is performed at a higher centre-of-mass energy ($8~\TeV$) than the previous measurements. Similar measurements have also been made previously in deep inelastic scattering and photoproduction experiments at HERA~\cite{Aaron:2007aa,Chekanov:2009dq,Aaron:2010uj,tagkey2014293}. 

\label{sec:intro_fid}

The fiducial region of the measurement presented is defined in terms of the photon kinematic quantities:\footnote{\AtlasCoordFootnote} 
transverse energy $\ETg$, pseudorapidity $\etag$ and transverse isolation energy $\ETiso$. 
The differential cross section is measured as a function of $\ETg$, for the highest-energy photon in the event, and spans the $25 < \ETg < 1500~\GeV$ range.
The $\etag$ range is split to give four intervals for the cross-section measurement: $|\etag| < 0.6$, $0.6 \leq |\etag| < 1.37$, $1.56 \leq |\etag| < 1.81$ and $1.81 \leq |\etag| < 2.37$.
The final constraint is the photon isolation, where $\ETiso$ is calculated within a cone of size $\Delta R = 0.4$, centred around the photon, and is chosen to be $\ETiso~<~4.8~\GeV~+~4.2~\times~10^{-3}~\times~\ETg$.
This fiducial region is identical in both the theoretical calculations and the experimental measurement; however, there are differences in the calculation of $\ETiso$:
\begin{itemize}
\item At detector level it is the sum of energy deposits in the calorimeter, corrected for the deposits related to the photon candidate itself.
\item At particle level it is the sum of energy from all particles, except for muons, neutrinos and the photon itself.
\item At parton level it is the sum of energy from all coloured partons.
\end{itemize}
An additional correction to remove energy from the underlying event (UE) or additional proton--proton interactions is applied at detector and particle level, as detailed in Section~\ref{sec:selection_isolation}. 

There are several differences between the measurement presented here and the previous ATLAS inclusive photon measurements~\cite{Aad:2010sp,Aad:2011tw,Aad:2013zba}. 
In addition to the change in centre-of-mass energy and $\ETg$ reach, it also probes for the first time the region 
$25< \ETg < 45~\GeV$ for $1.81 \leq |\etag| < 2.37$. The measurement is also compared to different theoretical predictions than used previously, as detailed in Section~\ref{sec:theory}.
An $\ETg$-dependent isolation requirement is introduced for the first time, effectively relaxing the maximum $\ETiso$ at high $\ETg$, as outlined in Section~\ref{sec:selection} along with the discussion of changing the upper edge of the excluded $\etag$ region from 1.52 to 1.56.
Other differences in the background estimation, unfolding and uncertainty calculations are highlighted in Sections~\ref{sec:background},~\ref{sec:cross_section} and ~\ref{sec:systematics} respectively, and the results are shown in Section~\ref{sec:result}.

\section{ATLAS detector and data}
\label{sec:detector}

\label{sec:atlas1}

The ATLAS experiment~\cite{Aad:2008zzm} at the LHC is a multi-purpose particle detector
with a forward-backward symmetric cylindrical geometry and a near $4\pi$ coverage in
solid angle. 
It consists of an inner tracking detector surrounded by a thin superconducting solenoid
providing a \SI{2}{\tesla} axial magnetic field, electromagnetic and hadronic calorimeters, and a muon spectrometer.
The inner tracking detector covers the pseudorapidity range $|\eta| < 2.5$.
It consists of silicon pixel, silicon microstrip, and transition radiation tracking detectors.
Within the region $|\eta|< 3.2$, EM calorimetry is provided by high-granularity lead/liquid-argon (LAr) sampling calorimeters,
with an additional thin LAr presampler covering $|\eta| < 1.8$,
to correct for energy loss in material upstream of the calorimeters.
A hadronic (steel/scintillator-tile) calorimeter covers the central pseudorapidity range ($|\eta| < 1.7$).
The end-cap and forward regions are instrumented with LAr calorimeters
for both the EM and hadronic energy measurements up to $|\eta| = 4.9$.
The muon spectrometer surrounds the calorimeters and is based on
three large air-core toroid superconducting magnets with eight coils each.
It includes a system of precision tracking chambers and fast detectors for triggering.
A three-level trigger system is used to select events.
The first-level trigger is implemented in hardware and uses a subset of the detector information
to reduce the accepted rate to at most \SI{75}{\kilo\hertz}.
This is followed by two software-based high-level triggers that
together reduce the accepted event rate to \SI{400}{\hertz} on average,
depending on the data-taking conditions during 2012.

The dataset used in this analysis was obtained using proton--proton collisions recorded in 2012 by the ATLAS detector, when the LHC operated at a centre-of-mass energy of $\sqrt{s} = 8~\TeV$.
The integrated luminosity of the dataset used in this measurement is 20.2~\ifb with
an uncertainty of 1.9\%~\cite{Aad:2013ucp}.
The events used in the analysis were recorded by the trigger system using single-photon triggers~\cite{ATLAS-CONF-2012-048}, which use identification criteria looser than the selection described in
Section~\ref{sec:selection_id}. For the high-level triggers, $\ETg$ thresholds are defined in
$20~\GeV$ steps from $20~\GeV$ to $120~\GeV$. Multiple trigger thresholds are required because the triggers are prescaled to reduce their
rate, except for the unprescaled $120~\GeV$ threshold. Each threshold is used in the analysis within an exclusive $\ETg$ range, determined to be where the trigger has an efficiency greater than 99.5\%, with respect to the full selection detailed in Section~\ref{sec:selection}. Only events taken during periods of good data
quality, where the calorimeters and inner tracking detectors are in nominal operation, are retained in the dataset. To remove any non-collision background, each event is
required to have a reconstructed vertex consistent with the average beam-spot position,
where the vertex is required to have at least two associated tracks. This condition is close to 100\% efficient for retaining events with photons within the detector acceptance.

\section{Theoretical predictions}
\label{sec:theory}

The theoretical calculations used in the analysis consist of LO Monte Carlo (MC)
event generators and calculations at next-to-leading-order (NLO) or higher.
Two event generators are used at LO: \PYTHIA~8.165~\cite{pythia} and \SHERPA~1.4.0~\cite{Gleisberg:2008ta}.
These event generators are interfaced with a detailed detector simulation~\cite{Aad:2010ah} (based on GEANT4~\cite{GEANT4}), the output of which is reconstructed in the same way as the data. The LO
predictions are used to study many aspects of the analysis and are also compared to the final cross section. The final cross sections are also compared to three calculations: \JETPHOX~\cite{jetphox}, \PETER~\cite{Becher:2012xr,Schwartz:2016olw} and \MCFM~\cite{Campbell:2010ff}.

Event generation with \PYTHIA\ includes: the description of the PDFs using CTEQ6L1~\cite{cteq}, the simulation of initial- and final-state radiation, the simulation of the UE using the ATLAS AU2 set of tuned parameters (tune)~\cite{ATL-PHYS-PUB-2012-003} based on the multiple parton interaction model~\cite{MPI}, and the modelling of the hadronisation based on the Lund string model~\cite{lund}. The LO direct contribution to the prompt photon production is fully included in the main matrix-element calculation. In contrast, the fragmentation contribution is modelled by final-state QED radiation arising from calculations of all $2\ra2$ QCD processes.

\PYTHIA\ is used to extract the central values of the measurement, while \SHERPA\ is used as a second LO generator as it showed excellent agreement with the results in the ATLAS photon plus jet measurement~\cite{Aad:2013gaa}. The \SHERPA\ predictions are used to cross-check the results and determine uncertainties arising from the use of MC simulations in parts of the analysis. The \SHERPA\ calculations are performed with up to four parton emissions and the radiation of gluons and photons is done coherently. This means that the fragmentation contribution is produced differently to the contribution in \PYTHIA\ and is also indistinguishable from the direct contribution, unlike \PYTHIA\ where the contributions can be separated.
The \SHERPA\ events are produced with: the CT10~\cite{Lai:2010vv} PDF,
the UE model based on the recommended tune provided by the \SHERPA\ authors, and hadronisation modelled using a modified version of the cluster model~\cite{ClusterModel}.

The LO simulated events used in the analysis are reweighted in order to match as well as possible the experimental conditions of the dataset. One of these corrections is to reproduce the pile-up (additional proton--proton interactions in the same bunch crossing) conditions, where the weights are derived from the distribution of average interactions per bunch crossing ($\mu$) in data and MC simulations
with an additional constant to improve the agreement of the number of primary vertices.  A second weight is used to ensure an accurate $\etag$ measurement by reproducing in the MC simulations the $z$-vertex position of the hard interaction measured in data.

The final cross sections are compared to these LO generators and also to parton-level calculations. The kinematic selection used in all of the predictions matches the fiducial region defined in Section~\ref{sec:intro_fid}. For the higher order predictions the nominal renormalisation ($\mu_{\mathrm R}$), factorisation ($\mu_{\mathrm F}$) and
fragmentation ($\mu_{\mathrm f}$) scales were set to the photon transverse
energy ($\mu_{\mathrm R} = \mu_{\mathrm F} = \mu_{\mathrm f} = \ETg$).

\JETPHOX, a well-established NLO parton-level generator for the prediction of processes with photons in the final state, is used as the baseline to compare the results. \JETPHOX\ is capable of calculating the double-differential inclusive prompt photon cross section $\mathrm{d}^{2}\sigma/(\mathrm{d}\ETg \mathrm{d}\etag)$ at parton level to NLO accuracy for both the direct and fragmentation photon
processes. The calculation can be configured to use an $\ETg$-dependent isolation requirement\footnote{The \ETiso\ requirement selected in this analysis is chosen to not be too restrictive for the NLO calculations, to avoid potential unphysical values in these predictions~\cite{jetphox}.} and uses the NLO photon fragmentation function of BFG set II~\cite{Bourhis:1998,Bourhis:2000gs}.
To check the effect of the PDF choice on the predictions, they are generated with different PDF sets (CT10, MSTW2008NLO~\cite{Martin:2009iq}, NNPDF2.3~\cite{Ball:2012cx} and HERAPDF1.5~\cite{herapdf}), provided by the
\LHAPDF\ package~\cite{Giele:2002hx}. The strong coupling constant ($\alphas$) is also obtained for each PDF using \LHAPDF\ and the fine-structure constant ($\alphaEM$) is set to the \JETPHOX\ default of $1/137$.

The following systematic uncertainties (combined in quadrature) are assigned to the \JETPHOX\ calculations and are estimated by means of procedures~\cite{Blair:1296265} used in the previous measurements:
\begin{itemize}
\item The uncertainty on the scale choice is evaluated from the envelope of varying the three scales by a factor of two around the nominal value, both simultaneously and independently (keeping two fixed at the nominal value). 
The impact on the predicted cross section varies between 12\% and 20\%.

\item The PDF uncertainty is obtained by repeating the \JETPHOX\ calculation for the
52 eigenvector sets of the CT10 PDF and applying a scaling factor in order to produce the
uncertainty for the 68\% confidence-level (CL) interval.
The corresponding uncertainty in the cross section increases with $\ETg$
and varies between 5\% at $100~\GeV$ and 15\% at $900~\GeV$.

\item The uncertainty due to $\alphas$ is evaluated,
following the recommendation of Ref.~\cite{Lai:2010vv},
by repeating the calculation with $\alphas$ varied by $\pm 0.002$ around
the central value of 0.118 and scaling in order to obtain the
uncertainty for the 68\% CL interval.
The uncertainty due to $\alphas$ is smaller than that from the scale or PDF uncertainties for the whole phase space; it slowly decreases from 9\% with increasing \ETg, with the exception of above $900~\GeV$ where it increases to 15\%.

\item To be able to correct from parton level to particle level, additional hadronisation-plus-UE correction factors were evaluated using the two alternative hadronisation and UE models in \PYTHIA\ and \SHERPA.
The study was performed by repeating the calculation with and without the hadronisation and UE contributions and resulted in a correction
close to unity for both MC models
with a small deviation of at most 2\% at low \ETg. Therefore, as in the previous analyses, no correction factor is applied to the central value; however, in this measurement an $\ETg$-dependent uncertainty is assigned to the theory, based on the largest deviation from unity between the two models.
\end{itemize}

\PETER\ is used as a second parton-level generator to predict the differential isolated prompt photon cross section at NLO including the resummation of threshold logarithms at the next-to-next-to-next-to-leading-logarithmic (NNNLL) level. \PETER\ is roughly equivalent to a fixed-order calculation at next-next-to-leading-order (NNLO); there is currently no exact calculation available for inclusive photons at this order. To account for the isolation criteria applied in the measurement, the \PETER\ result at NLO is normalised to that from \JETPHOX. The \PETER\ predictions are supplemented with the resummation of large electroweak Sudakov logarithms according to Ref.~\cite{Becher:2013zua,Becher:2015yea}. These electroweak corrections, not included in the predictions from \JETPHOX, provide estimates of electroweak uncertainties that are important for high $\ETg$ and also mean that, unlike \JETPHOX, \PETER\ uses a running $\alphaEM$. The scale uncertainty is calculated similarly to \JETPHOX, by varying the scales around the central value, but in \PETER\ there are four scales~\cite{Becher:2009th}: hard matching, jet, soft and factorisation. Finally the PDF uncertainty is taken directly from \JETPHOX.

An additional study was made using \MCFM, following on from the studies in Ref.~\cite{ATL-PHYS-PUB-2013-018}, with parameters (CT10 PDF, photon isolation, scale choice and $\alphaEM$) matching those in \JETPHOX. \MCFM\ calculates the fragmentation
process only to LO and therefore deviations from \JETPHOX\ predictions were expected below approximately $200~\GeV$. Surprisingly, however, even at higher \ETg\ the predictions from \MCFM\ were found to be consistently below the predictions from \JETPHOX, although within the theoretical uncertainties. This trend is under investigation by the calculations authors and the predictions are not presented here.

\section{Photon selection}
\label{sec:selection}

The photon selection, in both data and MC simulation, is based on the reconstruction~\cite{Aaboud:2016yuq} of an EM cluster in the calorimeter as a photon candidate.
The absence of an associated track in the inner detector classifies the photon candidate as an \textit{unconverted} photon, whereas it is classified as a \textit{converted} photon if the cluster is matched to two tracks coming from a conversion vertex or to one track which has no hits in the innermost layer of the inner tracking detector. Both the converted and unconverted candidates are kept in the analysis. A further track-based classification~\cite{ATL-PHYS-PUB-2011-007} is used to minimise the number of electrons reconstructed as photons, although this introduces a slight decrease in efficiency for reconstructing converted photons. The conversion classification is used both to determine the size of the photon cluster in the barrel calorimeter and also as an input to the dedicated energy calibration~\cite{Aad:2014nim}, which is applied to account for energy loss before the EM calorimeter. 
This calibration starts by correcting the response from each of the layers in the EM calorimeter and then applies a response calibration from MC simulations to the cluster energies. After accounting for detector response variations not included in the simulation, such as high-voltage inhomogeneities in some sectors, energy scale factors are then applied from the comparison of the detector response to $Z$ boson decays to electron--positron pair events in data and MC simulations.

Following this calibration, only photon candidates with $\ETg > 25~\GeV$ and a cluster barycentre (in the second layer of the EM calorimeter) lying within $|\etag| < 1.37$ or $1.56 \leq |\etag| < 2.37$ are retained for the analysis. The transition region between the barrel and end-cap calorimeters $(1.37 \leq |\etag| < 1.56)$ is excluded due to the degraded performance induced by the increased amount of inactive material in front of the calorimeter. This region is expanded in the measurement presented here to $1.56$, compared to the value of $1.52$ used previously, to improve the accuracy of the photon energy measurement as it avoids using clusters calibrated by scintillators that are part of the hadronic calorimeter.
Finally, photons reconstructed near regions of the calorimeter affected by read-out or high-voltage failures are not included in the analysis.
The remaining photon candidates are then used in this analysis if they satisfy further selection and quality criteria based on their calorimeter shower shapes and isolation energy.

\subsection{Photon identification}
\label{sec:selection_id}

In order to reduce the previously mentioned largest background, namely non-prompt photons originating mainly from decays of energetic $\pi^{0}$ and $\eta$ mesons, nine shower-shape variables~\cite{Aaboud:2016yuq} are exploited, similarly to the previous ATLAS inclusive photon measurements. These shower-shape variables are formed based on the relative and absolute energy deposition within the calorimeter cells using the full granularity of the different layers of the calorimeter system. The particular selection criteria for each of the nine variables are tuned for converted and unconverted photons separately, as well as being adjusted depending on $\etag$ (in intervals matching the four $\etag$ regions of this measurement). In the MC simulations the same criteria are applied as in data, but with two corrections. Firstly, the shower-shape variables are shifted~\cite{Aaboud:2016yuq} to match the measured distributions in data. Secondly, additional correction factors (at most a few percent from unity) to match the identification efficiency in the MC simulations and that in data are applied, calculated in each \ETg\ and $\etag$ interval. 

To quantify the effect of the identification criteria, the identification efficiency for prompt photons is defined in MC simulations as:
\begin{equation}
 \epsilon^{\mathrm{MC}}_{\mathrm{id}} = \frac{N^{\mathrm{MC}}_{\mathrm{id,matched}}}{N^{\mathrm{MC}}_{\mathrm{particle}}}.
\end{equation}
where reconstructed photons have to satisfy the identification criteria and be geometrically matched, with $\Delta R < 0.2$, to isolated photons generated at particle level. This $\epsilon^{\mathrm{MC}}_{\mathrm{id}}$ is shown in Figure~\ref{fig:efficiencies} along with the efficiencies for converted and unconverted photons.\footnote{At particle level the conversion classification is based on information from the detailed detector simulation of the photon, by searching for a conversion of the photon into an electron--positron pair within the geometrical region of the inner tracking detector.} The unconverted photon efficiency is high and approximately constant for more energetic photons, as expected since they should leave a more pronounced shower in the detector. However, a drop in efficiency is observed when combining with converted photons. 
The efficiency to reconstruct conversions decreases at high \ETg\ ($>150~\GeV$) where it becomes more difficult to separate the two tracks from the conversions. These very close-by tracks are more likely to fail the tighter selections, including a transition radiation requirement, applied to single-track conversion candidates.

\begin{figure}[htbp]
  \centering
    \includegraphics[scale=0.4]{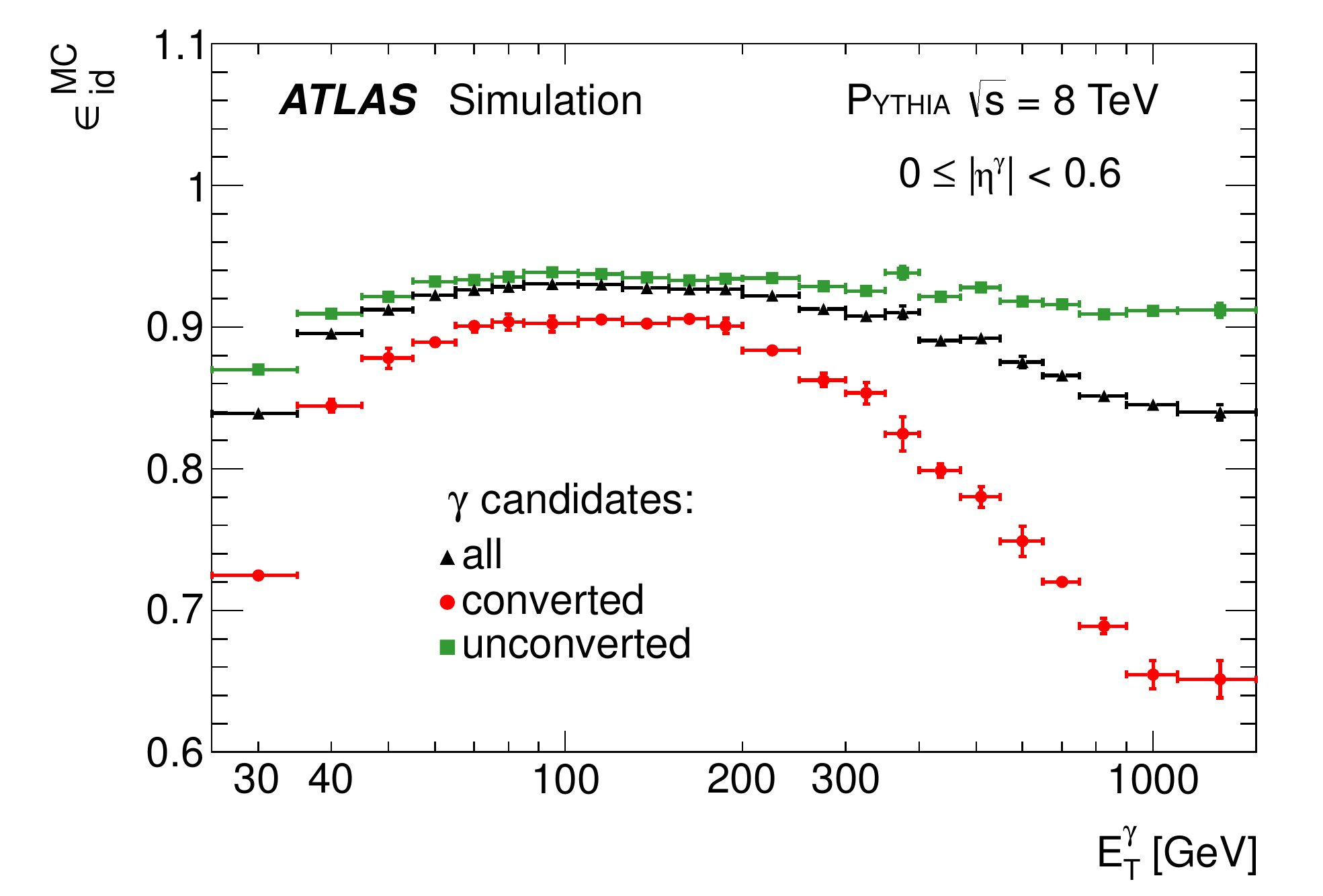}
  \caption{The photon identification efficiency (with statistical uncertainty) as a function of $\ETg$ determined in \PYTHIA\ MC simulations, along with the separated efficiencies for unconverted and converted photons. The efficiency is shown for the region $|\etag|<0.6$, with similar results found in other $|\etag|$ regions.}
  \label{fig:efficiencies}
\end{figure}

\subsection{Photon isolation}
\label{sec:selection_isolation}

The photon candidates are required to be isolated to distinguish between prompt photons and hadronic background. As stated in Section~\ref{sec:intro_fid}, $\ETiso$ is calculated from topological clusters of calorimeter cells in a cone of size $\Delta R = 0.4$ around the photon and corrected for the deposits related to the photon candidate itself.
As this quantity is susceptible to contributions from the UE and pile-up, a correction based on the jet area method~\cite{Cacciari:UE} is applied.
This estimates on an event-by-event basis the ambient energy density, which is then subtracted from the \ETiso\ before applying the isolation requirement. These corrections are typically between $1.5$ and $2~\GeV$. In order for the detector-level $\ETiso$ distribution to reproduce the distribution from data, it is corrected in each \ETg\ and $\etag$ interval by the difference between the mean value of $\ETiso$ in data and MC simulations. These corrections range from a few hundred \MeV\ up to 3--4~\GeV\ and are consistent for both \PYTHIA\ and \SHERPA.

The measurement presented here uses an $\ETg$-dependent isolation requirement:
\begin{equation}
\ETiso~<~4.8~\GeV~+~4.2~\times~10^{-3}~\times~\ETg.
\end{equation}
In contrast to the fixed value (3 or $7~\GeV$) used in the previous analyses, this requirement has been optimised to retain
more of the photons satisfying the identification criteria in Section~\ref{sec:selection_id} whilst also obtaining the best signal-to-background ratio throughout the large $\ETg$ range of the measurement.
In addition, the fraction of photon candidates that have satisfied the identification criteria and subsequently also satisfy the isolation requirement, stays high and constant. This is due to the isolation requirement being relaxed at higher $\ETg$, compared to using a fixed cut.

\section{Background subtraction}
\label{sec:background}

The number of events with a photon candidate ($N^{\gamma,\mathrm{data}}$) satisfying the kinematic, identification and isolation selection criteria, as detailed in Section~\ref{sec:selection}, has contributions from hadronic background and electrons. These contributions are removed statistically by techniques detailed below.

The hadronic background (from meson decays and jets) is removed by a data-driven technique, as done in the previous ATLAS analyses. This technique uses a two-dimensional sidebands method based on the isolation and identification criteria. For the identification, photons either satisfy the full criteria of all the shower-shape variables outlined in Section~\ref{sec:selection_id} or an orthogonal selection which aims to maximise the hadronic background. This orthogonal selection is achieved by inverting four variables related to the first layer of the EM calorimeter, which has cells with a very small width in $\eta$. For isolation, photons are either isolated as defined in Section~\ref{sec:selection_isolation} or non-isolated by having $\ETiso~>~7.8~\GeV~+~4.2~\times~10^{-3}~\times~\ETg$. The four regions are then defined in data to be:
\begin{itemize}
\item $N^{A,\mathrm{data}}$: photon candidates satisfying both the isolation and identification criteria, i.e. $N^{\gamma,\mathrm{data}}$.
\item $N^{B,\mathrm{data}}$: photon candidates that are non-isolated, but satisfy the identification criteria.
\item $N^{C,\mathrm{data}}$: photon candidates that only satisfy the orthogonal identification criteria but are isolated.
\item $N^{D,\mathrm{data}}$: photon candidates that only satisfy the orthogonal identification criteria and are non-isolated.
\end{itemize}

As defined above, there is a $3~\GeV$ separation between the non-isolated region and the isolated region. This separation is used to limit the number of particle-level signal photons that fall into the background regions. To quantify this effect, signal leakage fractions are calculated in MC simulations: 
\begin{equation}
f^{K,\mathrm{MC}} = \frac{N^{K,\mathrm{MC}}_{\mathrm{signal}}} {N^{A,\mathrm{MC}}_{\mathrm{signal}}},
\end{equation}
with $K=B,C,D$. These leakage fractions are found to be small and are calculated in \PYTHIA\ for the central value, with \SHERPA\ used as a cross-check.

The two-dimensional sidebands method assumes that the two chosen variables are independent for the background. The isolation and identification criteria are chosen to minimise any such dependence, but any deviation from this assumption can be accounted for by using MC simulations to calculate the ratio:
\begin{equation}
R_{\mathrm{bkg}}=\frac{N^{A,\mathrm{MC}}_{\mathrm{bkg}}\cdot N^{D,\mathrm{MC}}_{\mathrm{bkg}}}{N^{B,\mathrm{MC}}_{\mathrm{bkg}}\cdot N^{C,\mathrm{MC}}_{\mathrm{bkg}}},
\end{equation}
where $N^{K,\mathrm{MC}}_{\mathrm{bkg}}$ are the number of background events in each of the regions $K=A,B,C,D$. For the central value the assumption, confirmed in a control region, that they are independent ($R_{\mathrm{bkg}}=1$) is used; however, $R_{\mathrm{bkg}}$ is varied in Section~\ref{sec:systematics} to obtain the systematic uncertainty of any potential dependence.

The four sideband regions, signal leakage fractions and $R_{\mathrm{bkg}}$ are then used to solve for $N^{A,\mathrm{data}}_{\mathrm{signal}}$ via:
\begin{equation}
N^{A,\mathrm{data}}_{\mathrm{signal}} = N^{A,\mathrm{data}} - R_{\mathrm{bkg}}\cdot \left( (N^{B,\mathrm{data}}-f^{B,\mathrm{MC}}N^{A,\mathrm{data}}_{\mathrm{signal}})\cdot\frac{(N^{C,\mathrm{data}}-f^{C,\mathrm{MC}}N^{A,\mathrm{data}}_{\mathrm{signal}})}{(N^{D,\mathrm{data}}-f^{D,\mathrm{MC}}N^{A,\mathrm{data}}_{\mathrm{signal}})}\right).
\end{equation}
This solution is used in the cross-section measurement via the signal purity, which is defined as:
\begin{equation}
P_{\mathrm{signal}} = \frac{N^{A,\mathrm{data}}_{\mathrm{signal}}}{ N^{A,\mathrm{data}}}.
\end{equation}

In all four $\etag$ regions, $P_{\mathrm{signal}}$ is found to rise with $\ETg$ from 60\% at $25~\GeV$ to 100\% at around $300~\GeV$. 
In the highest $\ETg$ interval the method is inaccurate due to a lack of events in the background regions so here the central value of $P_{\mathrm{signal}}$ from the previous $\ETg$ interval is used in the cross-section calculation.

Finally, after the above subtraction a remaining background of fake photons from electrons is accounted for. 
As in previous measurements, this is estimated using MC simulations, scaled to the measured integrated luminosity in data, of $Z$ and $W$ boson decays to electrons. Reconstructed photons from these simulations passing the selection of Section~\ref{sec:selection} are counted if they are geometrically matched to a particle-level electron. 
The number of fake photons removed ($N_{e\ra\gamma}$) is less than 0.2\% of the remaining signal photons ($N^{\gamma,\mathrm{data}}P_{\mathrm{signal}}$) in all four $\etag$ regions and for most of the $\ETg$ range -- only reaching a maximum of 0.7\% in some low $\ETg$ intervals. As this is such a small effect no systematic uncertainty is assigned to this subtraction.

\section{Cross section}
\label{sec:cross_section}

The differential isolated prompt photon cross section as a function of \ETg\ (calculated in four $|\etag|$ regions) includes elements described in the previous sections and takes the form:
\begin{equation}
 \frac{\mathrm{d}\sigma}{\mathrm{d}\ETg} = 
 \frac{1}{\int \mathcal{L} \mathrm{d}t (\Delta{\ETg})} \;
 \cdot\; (
 N^{\gamma,\mathrm{data}}  \;
 \cdot\;
 P_{\mathrm{signal}} \;
 - \; 
 N_{e\ra\gamma}) \;
 \cdot\;
 \frac{1}{\epsilon_{\mathrm{trig}}} \;
 \cdot\;
 \frac{1}{\epsilon_{\mathrm{corr}}}, 
 \label{eqn:cross_section}
\end{equation}

where \ETg\ is that of the highest transverse energy photon satisfying the kinematic, identification and isolation criteria (Section~\ref{sec:selection}). The trigger efficiency ($\epsilon_{\mathrm{trig}}$) corrects $N^{\gamma,\mathrm{data}}$ for any events that would satisfy the selection criteria but were not recorded in the dataset (Section~\ref{sec:detector}). The number of events ($N^{\gamma,\mathrm{data}}$) with a photon satisfying the selection criteria is corrected for background using the previously introduced subtraction factors $P_{\mathrm{signal}}$ and 
 $N_{e\ra\gamma}$ (Section~\ref{sec:background}). Further, the overall size of the studied dataset is accounted for by dividing by the total integrated luminosity ($\int \mathcal{L} \mathrm{d}t$) and the cross section is normalised to inverse \GeV\ by dividing each measured \ETg\ interval by its size ($\Delta{\ETg}$). 
 
The remaining factor, $\epsilon_{\mathrm{corr}}$, is the unfolding correction factor 
used to correct the measurement to particle level 
to allow for direct comparisons to theoretical predictions. The unfolding factors are derived using \PYTHIA, with \SHERPA\ used as a cross-check. 
The unfolding correction factors are extracted by using a bin-by-bin unfolding procedure and are defined as:
\begin{equation}
 \epsilon_{\mathrm{corr}} = \frac{N_{\mathrm{signal}}^{\mathrm{MC}}}{N_{\mathrm{particle}}^{\mathrm{MC}}},
\end{equation} 
where $N_{\mathrm{signal}}^{\mathrm{MC}}$ and $N_{\mathrm{particle}}^{\mathrm{MC}}$ refer to the number of events with an isolated photon at detector level and particle level respectively.

The main contribution to $\epsilon_{\mathrm{corr}}$ is the identification efficiency (Section~\ref{sec:selection_id}), resulting in a very similar shape including the slight decrease at high \ETg. However, $\epsilon_{\mathrm{corr}}$ differs as it also contains the effects from photon migrations between different \ETg\ intervals and the isolation efficiency (Section~\ref{sec:selection_isolation}). The overall correction lies between 0.8 and 0.9 and therefore indicates that detector effects are rather small.   

The results of the bin-by-bin unfolding procedure are cross-checked using an iterative unfolding method, which reduces the reliance on the shape of the MC simulation distributions of $\ETg$ at particle or detector level. The method is based on Bayes Theorem~\cite{Bayes} and iteratively\footnote{In this analysis the result converges after four iterations.} unfolds the spectrum by changing the prior of the particle-level distribution to the previously unfolded spectrum for the next iteration. The results show that the two unfolding procedures are in very good agreement, considering statistical uncertainties only.

\section{Uncertainties}
\label{sec:systematics}

To estimate the systematic uncertainties, the cross-section calculation was repeated varying the selection procedure, background subtraction techniques or the unfolding correction factor. One difference compared to the previous analyses is that this measurement makes use of the Bootstrap technique~\cite{Bohm:2010tb} to evaluate the statistical influence on systematic uncertainties, achieved by producing a large number of weighted (based on a Poisson distribution) replicas for each event.
The result is then used to reduce the statistical fluctuations by applying a two-step smoothing technique; firstly
combining $\ETg$ intervals until the propagated uncertainty has a sufficiently large  statistical significance, followed by performing a Gaussian kernel smoothing on the original $\ETg$ intervals.

The following text describes the included uncertainty sources (quantifying those that are smaller):

\begin{itemize}

\item The photon energy scale is altered by varying systematic sources up and down, with the resulting shifts being summed in quadrature to provide the total uncertainty. The sources are split to account for correlations and range from being related to: detector material and read-out; simulation of the detector; extrapolations from data-driven measurements; and finally details related to the differences between unconverted or converted photon showers in the calorimeter. The uncertainty in the photon energy scale is around 1\%, except for the region $1.56 \leq |\etag| < 1.81$, but the uncertainty in the measurement is larger due to the steeply falling cross section.

\item The admixture of direct and fragmentation photons in a given $\ETg$ interval affects the calculation of both $P_{\mathrm{signal}}$ and $\epsilon_{\mathrm{corr}}$. Instead of using the default MC simulation fraction, a fit of the $\ETg$ distribution is performed in \PYTHIA\ to find the optimal admixture (as done in the recent photon plus jet paper~\cite{Aad:2013gaa}). The uncertainty is derived by comparing the results from this optimal admixture with the default \PYTHIA\ simulation. This replaces the systematic uncertainty obtained previously from an arbitrary removal or doubling of the fragmentation component.

\item $R_{\mathrm{bkg}}$ is set to unity when $P_{\mathrm{signal}}$ is calculated. As described in Section~\ref{sec:background}, this follows the assumption that there are no correlations between the isolation and identification criteria for the background. A test of this assumption is performed by subdividing the background-dominated region with an additional non-isolated criterion and then repeating the two-dimensional sidebands in background only regions. A 10\% difference from unity is found in this test, which is then applied to $R_{\mathrm{bkg}}$ to calculate the uncertainty.

\item As described in Section~\ref{sec:selection_id}, the photon identification efficiency in the MC simulations uses correction factors and the associated uncertainty in these alters the cross section by 0.5\% for most of the $\ETg$ range. In the lowest $\ETg$ intervals it reaches 2\% and above $550~\GeV$ it ranges from 1\% to 4\% (increasing with $\etag$).

\item For the above photon identification correction factors an extra uncertainty is required, obtained from MC simulations, to account for a small difference in the photon isolation requirement applied in this analysis from that used for the measurement of the photon identification efficiency. This impacts the cross section by 0.5\% but rises to 1\% for the highest $\ETg$ intervals.

\item The orthogonal identification selection in Section~\ref{sec:background} relies on inverting the selection criteria of four of the shower-shape variables. The uncertainty in this procedure is estimated by inverting either only two of these variables or by inverting an extra variable. A data-driven technique is used to disentangle this uncertainty from that already included in the $R_{\mathrm{bkg}}$ uncertainty above. The resulting uncertainty is 2\% for $\ETg<100~\GeV$ but quickly falls to zero for higher $\ETg$.

\item The isolation requirement used to define the background region in the $P_{\mathrm{signal}}$ calculation was altered so that the constant part of the requirement ($7.8~\GeV$) was varied by $\pm 1~\GeV$ (chosen as it is larger than any difference in the MC simulations between particle-level and detector-level isolation). The resulting uncertainty is less than 0.5\%.

\item The photon energy resolution is calculated from several independent sources in a similar manner to the energy scale, but the resolution is found to be of much less importance than the scale as it only produces an uncertainty of 0.5\%, which rises to 1\% for the highest $\ETg$ intervals.

\item The effect of unfolding is investigated by using a smooth function to reweight the MC simulations to match the data $\ETg$ distribution. Unfolding the data using this reweighted MC prediction gives a difference of less than 0.5\% compared to the nominal value.

\item The uncertainty in the correction factors from the choice of QCD-cascade and hadronisation model is derived from comparing \SHERPA\ with \PYTHIA. To avoid double counting the effects from the fragmentation contribution, the \PYTHIA\ simulation with the optimal admixture of direct and fragmentation photons is used again. The resulting uncertainty is 2\% at low $\ETg$ but quickly falls to zero as $\ETg$ increases.

\item The integrated luminosity has an uncertainty measured to be $\pm1.9$\%. It is derived, following the same methodology as that detailed in Ref.~\cite{Aad:2013ucp}, from a 
calibration of the luminosity scale derived from beam-separation scans performed in November 2012.

\item Other uncertainties were studied, but are not included in the systematic uncertainty as they were found to be negligible. Examples of these studies include: investigating the trigger efficiency (statistical uncertainties are $< 0.1\%$), pile-up (splitting the dataset by number of interactions per bunch crossing) and the MC simulation isolation shift (correcting the MC simulation by twice the fit accuracy).

\end{itemize}

The systematic uncertainties except for the luminosity uncertainty are combined. This is done by treating each of the sources as uncorrelated in each $\ETg$ interval. However, the sources are treated as correlated across different intervals in $\ETg$. This combination is shown in Figure~\ref{fig:uncertSummary} along with several of the main systematic uncertainties detailed above. The energy scale uncertainty dominates the high-$\ETg$ region, especially in the region $ 1.56 < |\etag| < 1.81$. At low $\ETg$ the uncertainties from the $R_{\mathrm{bkg}}$ variation and admixture of direct and fragmentation photons are of similar magnitude and dominate the uncertainty. In the $\ETg$ range 80--200~\GeV\ the main systematic uncertainties are of similar order and, in all but the region $ 1.56 < |\etag| < 1.81$, this leads to the luminosity uncertainty being larger than this combination of the other systematic uncertainties.

\begin{figure}[htbp]
  \centering
  \includegraphics[scale=0.8 , angle=0]{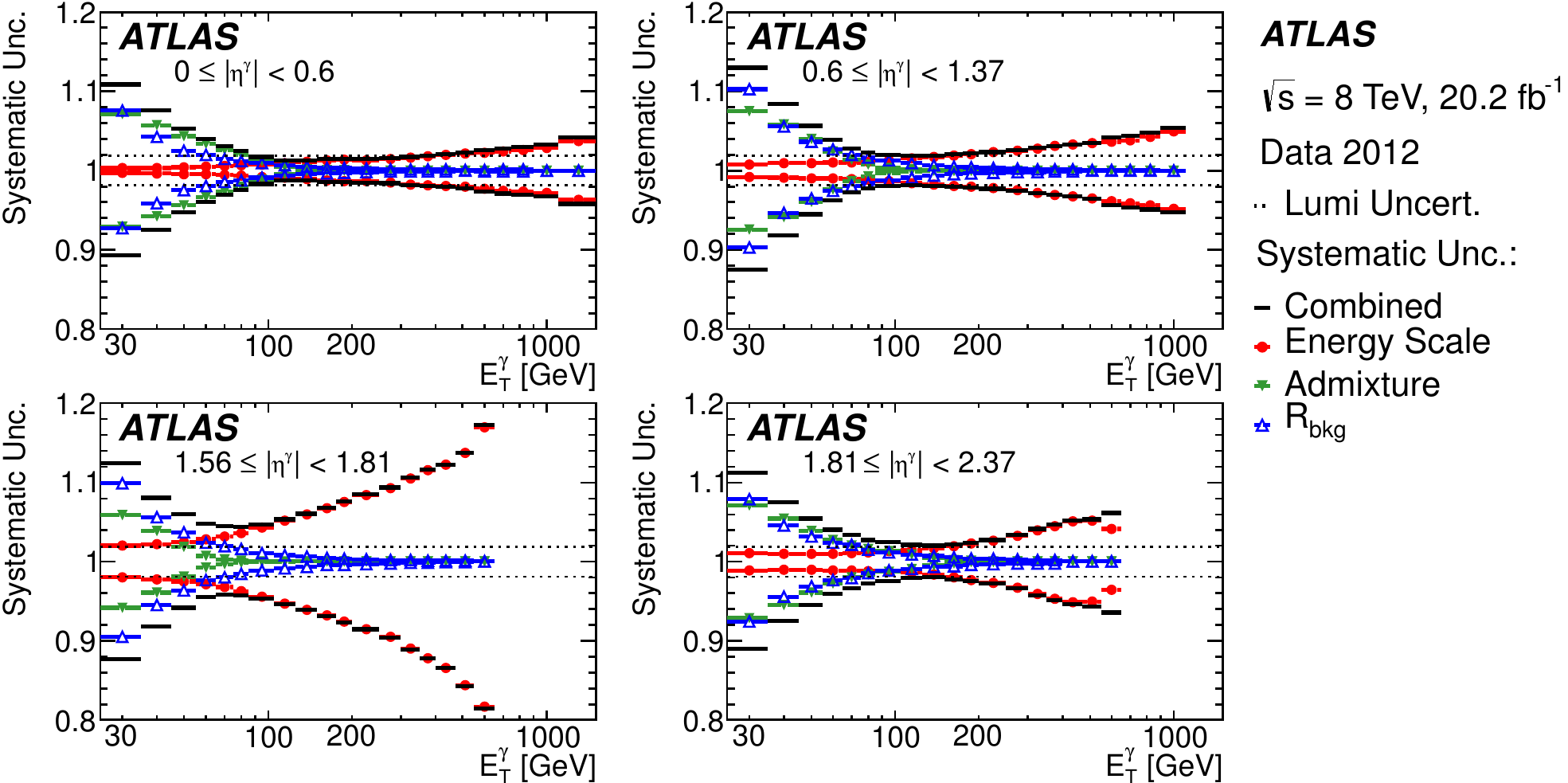}
\caption{Summary of the relative size of the combined systematic uncertainty (which excludes the luminosity) and its four main contributions, shown as a function of $\ETg$.
}
\label{fig:uncertSummary}
\end{figure}

The statistical uncertainty is mainly from the data, but also has a component due to the MC simulation. This component is from the reliance on MC simulations in the calculation of $P_{\mathrm{signal}}$ and $\epsilon_{\mathrm{corr}}$. The resulting total statistical uncertainty is 1--2\% for most of the measured $\ETg$ range, until it rises steeply in the highest $\ETg$ intervals.

\section{Results and discussion}
\label{sec:result}

The final cross sections are measured following Eqn.~\ref{eqn:cross_section} in the fiducial region given in Section~\ref{sec:intro_fid}. The systematic uncertainties, as described in Section~\ref{sec:systematics}, are combined with the statistical uncertainty, but do not include the luminosity uncertainty. The measured cross sections are compared to theoretical predictions, as detailed in Section~\ref{sec:theory}, along with uncertainties from the combination of the scale, PDF, $\alphas$ and hadronisation plus UE uncertainties. Figure~\ref{fig:xSec} shows a summary of the results (with the measured cross sections also being tabulated in Appendix~\ref{sec:app_tables}), where it can be seen that the measurement is well described overall by \JETPHOX\ over ten orders of magnitude in cross section. The total cross sections shown in Table~\ref{tab:totalcsjetphox} are integrated over the entire $\ETg$ for each $\etag$ region. As seen in the previous measurement~\cite{Aad:2013zba} the total cross sections are 20\% higher in data than those predicted by \JETPHOX, but the results are consistent within the uncertainties. It can also be seen that the measurement uncertainty, dominated by the systematic uncertainty, is smaller than the theoretical uncertainty.

\begin{table}[htbp]
	\begin{center}
    \renewcommand{\arraystretch}{1.4}
	\begin{tabular}{c c c c}
	\toprule
	$|\etag|$ range & \ETg\ range [\GeV] & Measured total $\sigma$ [nb] & \JETPHOX\ total $\sigma$ [nb] \\
	\midrule
	\pho\phantom{.}0--0.6 & 25--1500 &
$15.6^{+1.4}_{-1.4}(\mathrm{syst})\pm0.02(\mathrm{stat})\pm0.3(\mathrm{lumi})$ & 
$13.3\pm2.6(\mathrm{theory})$ \\
	\pho0.6--1.37 & 25--1100 &
$20.2^{+2.2}_{-2.1}(\mathrm{syst})\pm0.03(\mathrm{stat})\pm0.4(\mathrm{lumi})$ & 
$17.1\pm3.2(\mathrm{theory})$ \\
	1.56--1.81 & 25--650\pho &
$\pho 6.7^{+0.7}_{-0.7}(\mathrm{syst})\pm0.02(\mathrm{stat})\pm0.1(\mathrm{lumi})$ & 
$\pho 5.2\pm1.0(\mathrm{theory})$ \\
	1.81--2.37 & 25--650\pho &
$14.3^{+1.4}_{-1.3}(\mathrm{syst})\pm0.03(\mathrm{stat})\pm0.3(\mathrm{lumi})$ & 
$11.4\pm2.3(\mathrm{theory})$ \\
	\bottomrule
 	\end{tabular}
	\caption{Measured and predicted total cross sections shown for each of the four $|\etag|$ ranges. The \JETPHOX\ predictions are made using the CT10 PDF.}
	\label{tab:totalcsjetphox}
	\end{center}
\end{table}

\begin{figure}[htbp]
  \centering
  \includegraphics[scale=0.8 , angle=0]{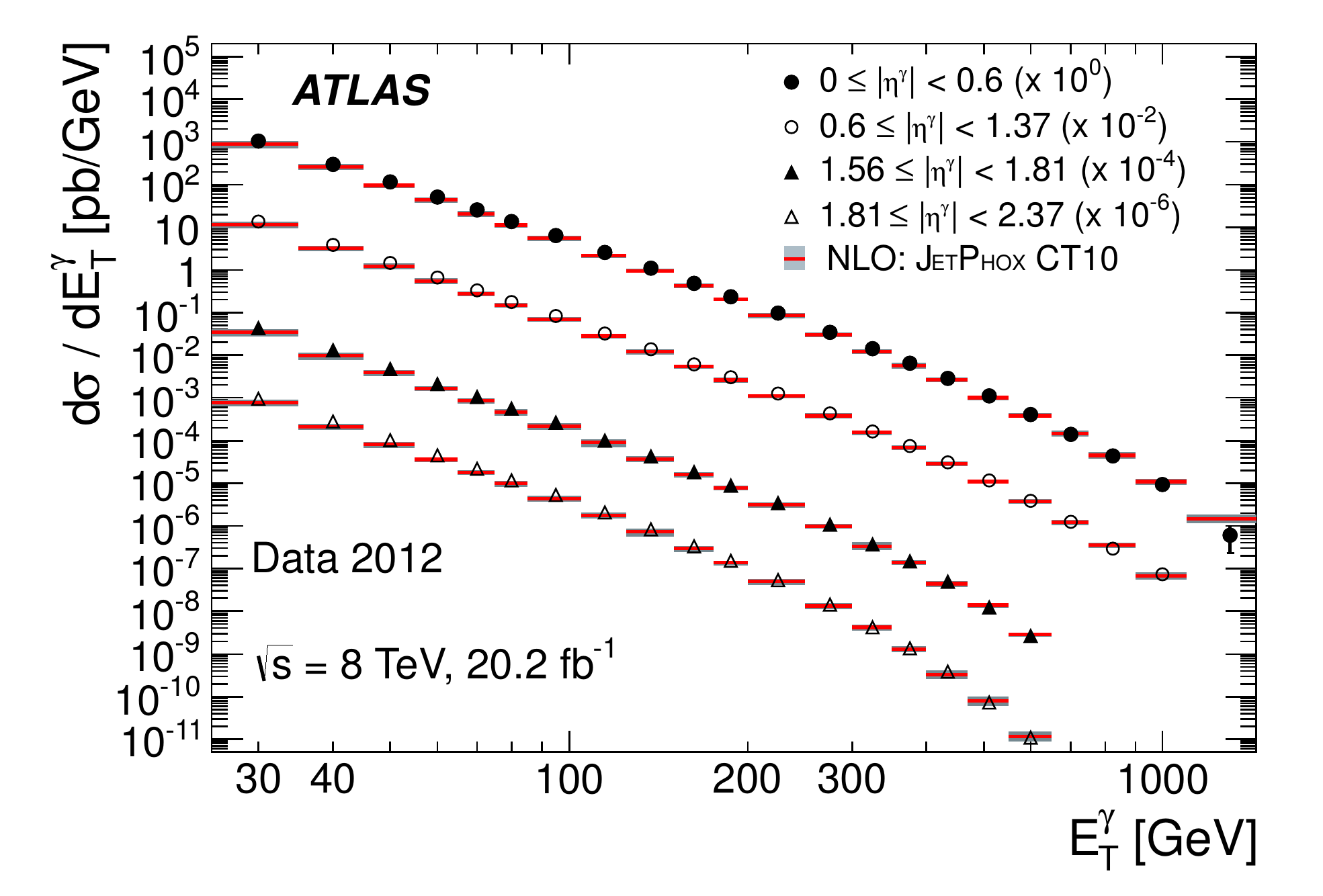}
\caption{Differential cross sections from data and \JETPHOX\ (using the CT10 PDF), shown as a function of $\ETg$ for the four $|\etag|$ regions. The distributions are scaled, by specified factors, to separate the distributions visually.}
\label{fig:xSec}
\end{figure}

The difference between data and \JETPHOX\ is explored further in Figure~\ref{fig:xSecRatio} where the cross-section ratios are shown in each of the four $\etag$ regions as a function $\ETg$. Each $\etag$ region shows a similar trend at low $\ETg$, in that the \JETPHOX\ NLO predictions are up to 20\% lower than those measured. This difference remains constant, especially in the central $\etag$ region, for $\ETg < 500~\GeV$ where the fragmentation contribution decreases with $\ETg$ from being a large contribution to the cross section, showing that \JETPHOX\ models this contribution well apart from the normalisation. The normalisation difference decreases above this $\ETg$ and in the range $1100 \leq \ETg < 1500~\GeV$ the prediction overestimates the measurement, although this is where the experimental and PDF uncertainties are largest. The results are shown using the CT10 PDF, but there is very little difference when comparing the central value to those from MSTW2008, NNPDF2.3 and HeraPDF1.5, with any difference at high $\ETg$ being covered by the large theoretical uncertainty. 

The overall trend in differences between data and theory is similar to that seen in the measurement using 2011 data. However, a significant increase in the experimental precision of this measurement compared to the previous ATLAS measurements reveals new qualitative features in the comparison to \JETPHOX. While the theoretical uncertainties have not changed, the measurement uncertainties are halved over most of the phase space.\footnote{Only in the region $1.56 \leq |\etag| < 1.81$ is the 2011 uncertainty comparable, as it is measured in a larger $\etag$ region.} This makes the uncertainties considerably smaller than the theoretical uncertainties, except in the statistically limited highest $\ETg$ intervals, which leads to disagreement in some $\ETg$ intervals between the measurement and the \JETPHOX\ prediction. This improvement in accuracy can help to reduce PDF uncertainties once the measurement is included in a global fit.

\begin{figure}[htbp]
  \centering
  \includegraphics[scale=0.8 , angle=0]{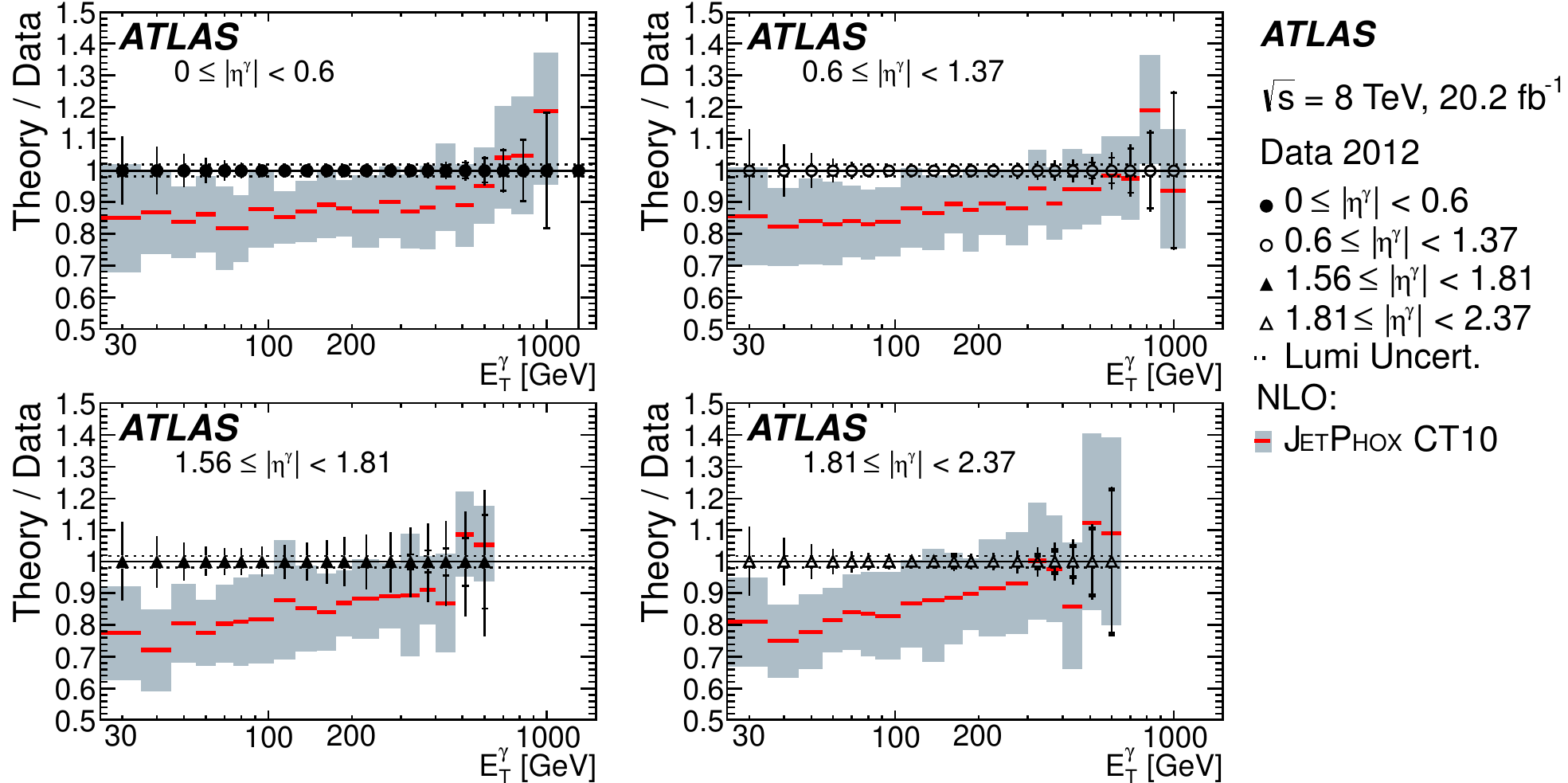}
\caption{Ratio of theory (\JETPHOX\ using the CT10 PDF) to data for the differential cross sections as a function of $\ETg$ for the four $|\etag|$ regions. The statistical component of the uncertainty in the data is indicated by the horizontal tick marks whereas the whole error bar corresponds to the combined statistical and systematic uncertainty (the additional systematic uncertainty arising from the uncertainty in the integrated luminosity is displayed separately as a dotted line). The NLO total uncertainty from \JETPHOX\ is displayed as a band, which corresponds to the combination of the scale, $\alphas$, PDF and hadronisation-plus-UE uncertainties. In the highest $\ETg$ interval of the $|\etag| < 0.6$ region the theoretical prediction and uncertainty is not shown as it is above the range of the figure.
}
\label{fig:xSecRatio}
\end{figure}

In order for the data to provide a tighter constraint on proton PDF uncertainties, it would be preferable both to have a better general agreement between data and the predictions and also to reduce the dominant theoretical scale uncertainties. This can be achieved by using calculations beyond NLO, as done here by using the predictions from \PETER. This comparison is shown in Figure~\ref{fig:xSecRatioPeter} where it can be seen that \PETER\ does an excellent job of removing the normalisation difference seen between data and \JETPHOX, especially in the region $|\etag| < 1.37$. The uncertainties shown, from combining the scale, PDF and electroweak uncertainties, are about 20\% lower than those from \JETPHOX. The \PETER\ predictions match the data well, within the combined measured and theoretical uncertainties, in all of the measured phase space. The improved normalisation and smaller uncertainties are also seen in the total cross sections as shown in Table~\ref{tab:totalcspeter}.

\begin{figure}[htbp]
  \centering
  \includegraphics[scale=0.8 , angle=0]{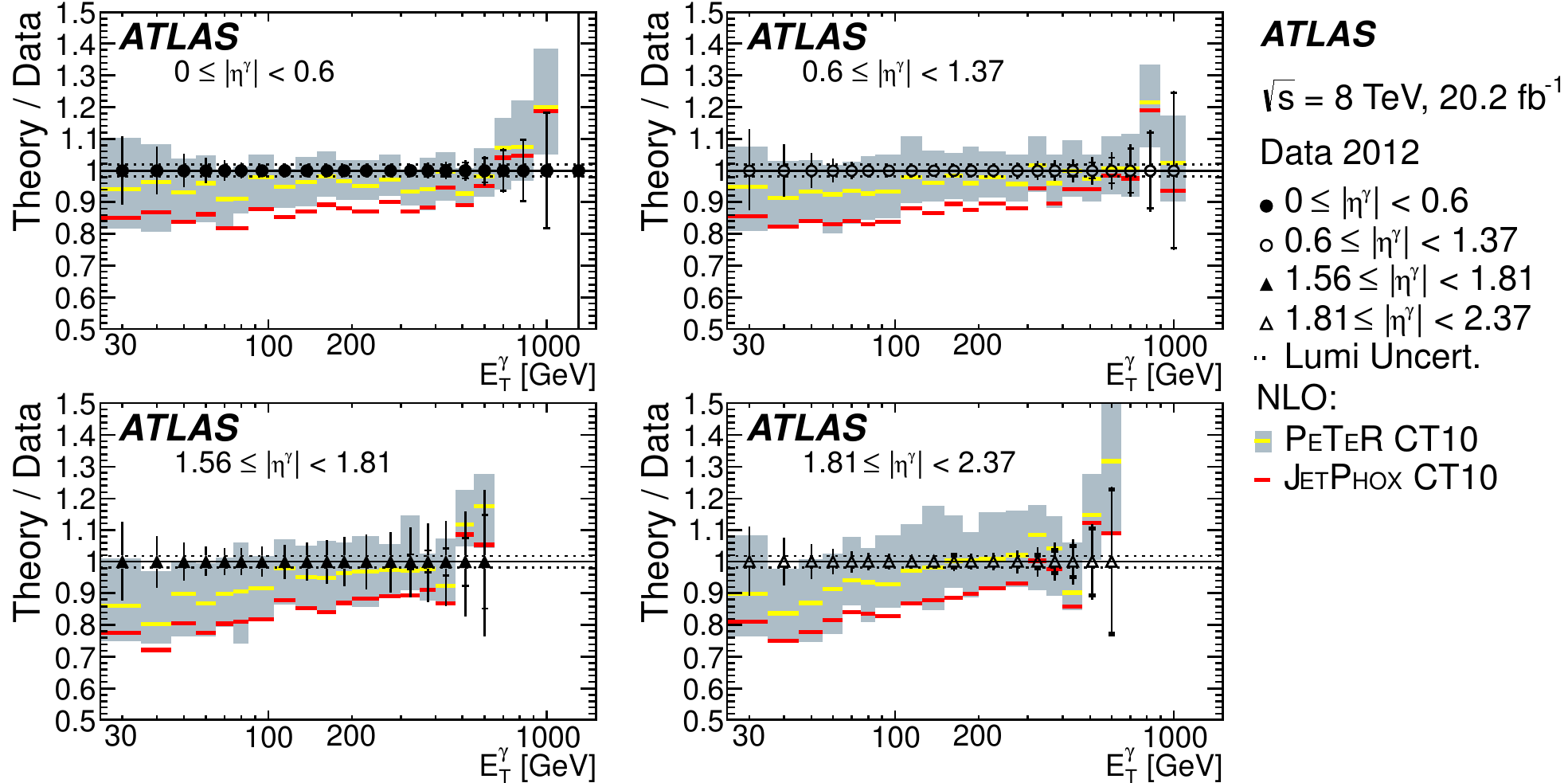}
\caption{Ratio of theory (\PETER\ and \JETPHOX\ both using the CT10 PDF) to data for the differential cross sections as a function of $\ETg$ for the four $|\etag|$ regions. The statistical component of the uncertainty in the data is indicated by the horizontal tick marks whereas the whole error bar corresponds to the combined statistical and systematic uncertainty (the additional systematic uncertainty arising from the uncertainty in the integrated luminosity is displayed separately as a dotted line). The NLO total uncertainty from \PETER\ is displayed as a band, which corresponds to the combination of the scale, PDF and electroweak uncertainties. In the highest $\ETg$ interval of the $|\etag| < 0.6$ region the theoretical predictions and uncertainty are not shown as they are above the range of the figure.
}
\label{fig:xSecRatioPeter}
\end{figure}

\begin{table}[htbp]
	\begin{center}
    \renewcommand{\arraystretch}{1.4}
	\begin{tabular}{c c c}
	\toprule
	$|\etag|$ range & \ETg\ range [\GeV] 
	& \PETER\ total $\sigma$ [nb] \\
	\midrule
	\pho\phantom{.}0--0.6 & 25--1500 &
$14.8\pm2.0(\mathrm{theory})$ \\
	\pho0.6--1.37 & 25--1100 &
$19.0\pm2.5(\mathrm{theory})$ \\
	1.56--1.81 & 25--650\pho &
$\pho 5.7\pm0.7(\mathrm{theory})$ \\
	1.81--2.37 & 25--650\pho &
$12.7\pm1.8(\mathrm{theory})$ \\
	\bottomrule
 	\end{tabular}
	\caption{Predicted total cross sections from \PETER\ shown for each of the four $|\etag|$ ranges, made using the CT10 PDF. }
	\label{tab:totalcspeter}
	\end{center}
\end{table}

Finally, the measured cross sections are also compared to the LO parton shower MC calculations in Figure~\ref{fig:xSecRatioLO}. Here it can be seen that generally \SHERPA, without any normalisation scaling, matches the data in the range $100 \leq \ETg < 500~\GeV$ in all four $\etag$ regions. At low $\ETg$, where a larger fragmentation contribution is expected, \SHERPA\ matches the predictions from \JETPHOX and thus is in disagreement with the measurement. At high $\ETg$ the \SHERPA\ prediction tends to be above the measured value. \PYTHIA\ on the other hand is similar to \JETPHOX for $\ETg > 100~\GeV$ and hence is below the measured cross section in all $\etag$ regions except $1.81 \leq |\etag| < 2.37$. At low $\ETg$, the \PYTHIA\ prediction has a very different shape than both the measurement and the other predictions, tending to overestimate the measured cross section, which suggests that the fragmentation contribution is not well modelled by the parton shower.

\begin{figure}[htbp]
  \centering
  \includegraphics[scale=0.8 , angle=0]{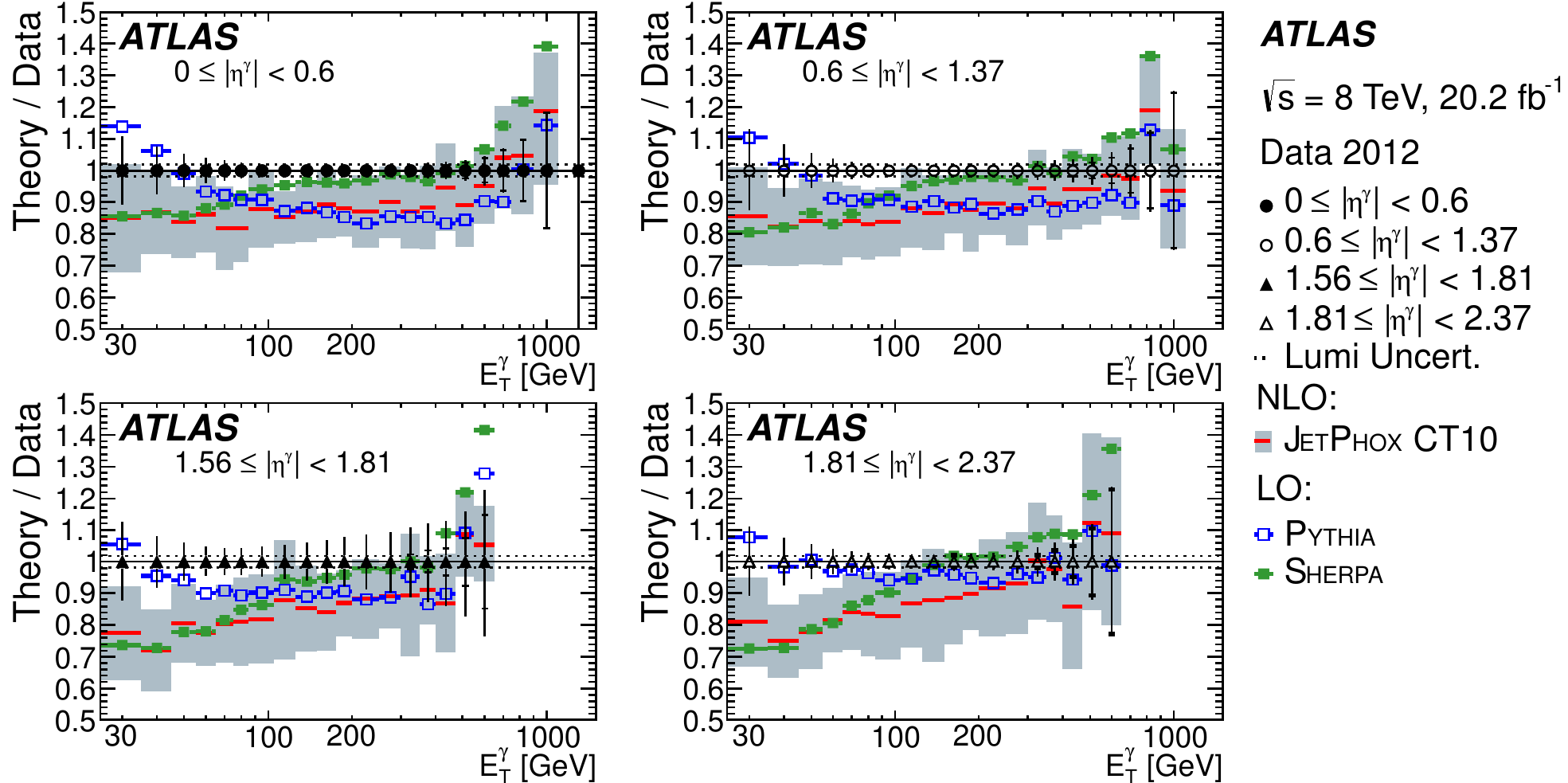}
\caption{Ratio of theory (\PYTHIA, \SHERPA\ and \JETPHOX) to data for the differential cross sections as a function of $\ETg$ for the four $|\etag|$ regions. The statistical component of the uncertainty in the data is indicated by the horizontal tick marks whereas the whole error bar corresponds to the combined statistical and systematic uncertainty (the additional systematic uncertainty arising from the uncertainty in the integrated luminosity is displayed separately as a dotted line). The NLO total uncertainty from \JETPHOX\ is displayed as a band, which corresponds to the combination of the scale, $\alphas$, PDF and hadronisation-plus-UE uncertainties. In the highest $\ETg$ interval of the $|\etag| < 0.6$ region the theoretical predictions and uncertainty are not shown as they are above the range of the figure.
}
\label{fig:xSecRatioLO}
\end{figure}


\FloatBarrier

\section{Conclusion}
\label{sec:conclusion}

In conclusion, a measurement of the inclusive isolated photon cross section has been presented, using 20.2~\ifb of $\sqrt{s} = 8~\TeV$ proton--proton collision data recorded by the ATLAS detector at the LHC. This is measured for the highest-energy photon in the event, spanning $25 < \ETg < 1500~\GeV$, in one of four $\etag$ regions ($|\etag| < 0.6$, $0.6 \leq |\etag| < 1.37$, $1.56 \leq |\etag| < 1.81$ and $1.81 \leq |\etag| < 2.37$) and with the isolation requirement $\ETiso~<~4.8~\GeV~+~4.2~\times~10^{-3}~\times~\ETg$ calculated within a cone of size $\Delta R = 0.4$. The results presented cover ten orders of magnitude in cross section, extending the measurement above $1~\TeV$ whilst also revisiting lower-$\ETg$ data points. The results show a significant improvement in experimental uncertainties over the previous measurements. The results are compared to \JETPHOX\ predictions, which, for most of the $\ETg$ range, have a similar shape but lie below the data. The predictions from \PETER\ agree much better in normalisation and, unlike \JETPHOX, are within the uncertainties of the measured cross section for the entire phase space measured, showing the need for higher-order calculations to better understand this process theoretically. Comparing the results to LO parton shower MC calculations shows different trends, with the largest differences being at low $\ETg$ in the region dominated by the fragmentation contribution. Finally, halving the measured uncertainties compared to previous measurements will make this a useful constraint on proton PDF uncertainties once the result is included in a global fit.

\section*{Acknowledgements}


We thank CERN for the very successful operation of the LHC, as well as the
support staff from our institutions without whom ATLAS could not be
operated efficiently.

We acknowledge the support of ANPCyT, Argentina; YerPhI, Armenia; ARC, Australia; BMWFW and FWF, Austria; ANAS, Azerbaijan; SSTC, Belarus; CNPq and FAPESP, Brazil; NSERC, NRC and CFI, Canada; CERN; CONICYT, Chile; CAS, MOST and NSFC, China; COLCIENCIAS, Colombia; MSMT CR, MPO CR and VSC CR, Czech Republic; DNRF and DNSRC, Denmark; IN2P3-CNRS, CEA-DSM/IRFU, France; GNSF, Georgia; BMBF, HGF, and MPG, Germany; GSRT, Greece; RGC, Hong Kong SAR, China; ISF, I-CORE and Benoziyo Center, Israel; INFN, Italy; MEXT and JSPS, Japan; CNRST, Morocco; FOM and NWO, Netherlands; RCN, Norway; MNiSW and NCN, Poland; FCT, Portugal; MNE/IFA, Romania; MES of Russia and NRC KI, Russian Federation; JINR; MESTD, Serbia; MSSR, Slovakia; ARRS and MIZ\v{S}, Slovenia; DST/NRF, South Africa; MINECO, Spain; SRC and Wallenberg Foundation, Sweden; SERI, SNSF and Cantons of Bern and Geneva, Switzerland; MOST, Taiwan; TAEK, Turkey; STFC, United Kingdom; DOE and NSF, United States of America. In addition, individual groups and members have received support from BCKDF, the Canada Council, CANARIE, CRC, Compute Canada, FQRNT, and the Ontario Innovation Trust, Canada; EPLANET, ERC, FP7, Horizon 2020 and Marie Sk{\l}odowska-Curie Actions, European Union; Investissements d'Avenir Labex and Idex, ANR, R{\'e}gion Auvergne and Fondation Partager le Savoir, France; DFG and AvH Foundation, Germany; Herakleitos, Thales and Aristeia programmes co-financed by EU-ESF and the Greek NSRF; BSF, GIF and Minerva, Israel; BRF, Norway; Generalitat de Catalunya, Generalitat Valenciana, Spain; the Royal Society and Leverhulme Trust, United Kingdom.

The crucial computing support from all WLCG partners is acknowledged
gratefully, in particular from CERN and the ATLAS Tier-1 facilities at
TRIUMF (Canada), NDGF (Denmark, Norway, Sweden), CC-IN2P3 (France),
KIT/GridKA (Germany), INFN-CNAF (Italy), NL-T1 (Netherlands), PIC (Spain),
ASGC (Taiwan), RAL (UK) and BNL (USA) and in the Tier-2 facilities
worldwide.


\clearpage
\appendix
\part*{Appendix}
\addcontentsline{toc}{part}{Appendix}

\section{Tables of measured cross sections}
\label{sec:app_tables}

The measured $\ETg$-differential cross sections are listed in 
Tables~\ref{tab:inclusivePhoton_first}, \ref{tab:inclusivePhoton_second}, \ref{tab:inclusivePhoton_third} and \ref{tab:inclusivePhoton_fourth}.

\begin{table}[ht]
\centering
   {\small
   \renewcommand{\arraystretch}{1.3}
\begin{tabular}{|c|ccccc|}
\hline 
$\ETg$ range [\GeV] &  $\mathrm{d}\sigma^{\gamma} / \mathrm{d} \ETg$ & Stat. Unc. & Sys. Unc. & Lumi. Unc. & [pb/\GeV] \\ \hline 
25--35   & $1.03$ 	 & $\pm 0.00$ 	 & $\pm^{0.11}_{0.11}$ 	& $\pm 0.02$ 	 & $\cdot 10^{3}$\\ \hline 
35--45   & $3.01$ 	 & $\pm 0.01$ 	 & $\pm^{0.23}_{0.22}$ 	& $\pm 0.06$ 	 & $\cdot 10^{2}$\\ \hline 
45--55   & $1.15$ 	 & $\pm 0.01$ 	 & $\pm^{0.06}_{0.06}$ 	& $\pm 0.02$ 	 & $\cdot 10^{2}$\\ \hline 
55--65   & $5.03$ 	 & $\pm 0.02$ 	 & $\pm^{0.20}_{0.20}$ 	& $\pm 0.10$ 	 & $\cdot 10^{1}$\\ \hline 
65--75   & $2.54$ 	 & $\pm 0.01$ 	 & $\pm^{0.08}_{0.08}$ 	& $\pm 0.05$ 	 & $\cdot 10^{1}$\\ \hline 
75--85   & $1.37$ 	 & $\pm 0.01$ 	 & $\pm^{0.03}_{0.03}$ 	& $\pm 0.03$ 	 & $\cdot 10^{1}$\\ \hline 
\pho85--105  & $6.37$ 	 & $\pm 0.03$ 	 & $\pm^{0.11}_{0.11}$ 	& $\pm 0.12$ 	 & \\ \hline 
105--125 & $2.54$ 	 & $\pm 0.01$ 	 & $\pm^{0.03}_{0.03}$ 	& $\pm 0.05$ 	 & \\ \hline 
125--150 & $1.09$ 	 & $\pm 0.00$ 	 & $\pm^{0.01}_{0.01}$ 	& $\pm 0.02$ 	 & \\ \hline 
150--175 & $4.84$ 	 & $\pm 0.02$ 	 & $\pm^{0.07}_{0.07}$ 	& $\pm 0.09$ 	 & $\cdot 10^{-1}$\\ \hline 
175--200 & $2.34$ 	 & $\pm 0.01$ 	 & $\pm^{0.03}_{0.03}$ 	& $\pm 0.04$ 	 & $\cdot 10^{-1}$\\ \hline 
200--250 & $9.84$ 	 & $\pm 0.05$ 	 & $\pm^{0.14}_{0.15}$ 	& $\pm 0.19$ 	 & $\cdot 10^{-2}$\\ \hline 
250--300 & $3.42$ 	 & $\pm 0.02$ 	 & $\pm^{0.05}_{0.06}$ 	& $\pm 0.07$ 	 & $\cdot 10^{-2}$\\ \hline 
300--350 & $1.41$ 	 & $\pm 0.01$ 	 & $\pm^{0.02}_{0.03}$ 	& $\pm 0.03$ 	 & $\cdot 10^{-2}$\\ \hline 
350--400 & $6.56$ 	 & $\pm 0.10$ 	 & $\pm^{0.13}_{0.13}$ 	& $\pm 0.13$ 	 & $\cdot 10^{-3}$\\ \hline 
400--470 & $2.84$ 	 & $\pm 0.05$ 	 & $\pm^{0.06}_{0.06}$ 	& $\pm 0.05$ 	 & $\cdot 10^{-3}$\\ \hline 
470--550 & $1.13$ 	 & $\pm 0.03$ 	 & $\pm^{0.03}_{0.02}$ 	& $\pm 0.02$ 	 & $\cdot 10^{-3}$\\ \hline 
550--650 & $4.05$ 	 & $\pm 0.15$ 	 & $\pm^{0.10}_{0.11}$ 	& $\pm 0.08$ 	 & $\cdot 10^{-4}$\\ \hline 
650--750 & $1.39$ 	 & $\pm 0.09$ 	 & $\pm^{0.04}_{0.04}$ 	& $\pm 0.03$ 	 & $\cdot 10^{-4}$\\ \hline 
750--900 & $4.36$ 	 & $\pm 0.42$ 	 & $\pm^{0.13}_{0.13}$ 	& $\pm 0.08$ 	 & $\cdot 10^{-5}$\\ \hline 
\pho900--1100 & $9.35$ 	 & $\pm 1.71$ 	 & $\pm^{0.31}_{0.31}$ 	& $\pm 0.18$ 	 & $\cdot 10^{-6}$\\ \hline 
1100--1500 & $6.12$ 	 & $\pm 3.80$ 	 & $\pm^{0.25}_{0.26}$ 	& $\pm 0.12$ 	 & $\cdot 10^{-7}$\\ \hline 
\end{tabular}

   }
\caption{
The inclusive prompt photon cross section with systematic and statistical uncertainties for the region $|\etag| < 0.6$.
}
\label{tab:inclusivePhoton_first}
\end{table}

\begin{table}[ht]
\centering
   {\small
   \renewcommand{\arraystretch}{1.3}
\begin{tabular}{|c|ccccc|}
\hline 
$\ETg$ range [\GeV] &  $\mathrm{d}\sigma^{\gamma} / \mathrm{d} \ETg$ & Stat. Unc. & Sys. Unc. & Lumi. Unc.& [pb/\GeV] \\ \hline 
25--35   & $1.34$ 	 & $\pm 0.00$ 	 & $\pm^{0.17}_{0.17}$& $\pm 0.03$ 	  	& $\cdot 10^{3}$\\ \hline 
35--45   & $3.88$ 	 & $\pm 0.01$ 	 & $\pm^{0.33}_{0.31}$& $\pm 0.07$ 	  	& $\cdot 10^{2}$\\ \hline 
45--55   & $1.44$ 	 & $\pm 0.01$ 	 & $\pm^{0.08}_{0.08}$& $\pm 0.03$ 	  	& $\cdot 10^{2}$\\ \hline 
55--65   & $6.61$ 	 & $\pm 0.03$ 	 & $\pm^{0.25}_{0.25}$& $\pm 0.13$ 	  	& $\cdot 10^{1}$\\ \hline 
65--75   & $3.30$ 	 & $\pm 0.01$ 	 & $\pm^{0.09}_{0.09}$& $\pm 0.06$ 	  	& $\cdot 10^{1}$\\ \hline 
75--85   & $1.77$ 	 & $\pm 0.01$ 	 & $\pm^{0.04}_{0.04}$& $\pm 0.03$ 	  	& $\cdot 10^{1}$\\ \hline 
\pho85--105  & $8.20$ 	 & $\pm 0.03$ 	 & $\pm^{0.16}_{0.16}$& $\pm 0.16$ 	  	& \\ \hline 
105--125 & $3.24$ 	 & $\pm 0.01$ 	 & $\pm^{0.06}_{0.06}$& $\pm 0.06$ 	  	& \\ \hline 
125--150 & $1.39$ 	 & $\pm 0.00$ 	 & $\pm^{0.03}_{0.03}$& $\pm 0.03$ 	  	& \\ \hline 
150--175 & $6.16$ 	 & $\pm 0.02$ 	 & $\pm^{0.13}_{0.13}$& $\pm 0.12$ 	  	& $\cdot 10^{-1}$\\ \hline 
175--200 & $3.01$ 	 & $\pm 0.02$ 	 & $\pm^{0.07}_{0.06}$& $\pm 0.06$ 	  	& $\cdot 10^{-1}$\\ \hline 
200--250 & $1.25$ 	 & $\pm 0.05$ 	 & $\pm^{0.03}_{0.03}$& $\pm 0.02$ 	  	& $\cdot 10^{-1}$\\ \hline 
250--300 & $4.31$ 	 & $\pm 0.03$ 	 & $\pm^{0.11}_{0.11}$& $\pm 0.08$ 	  	& $\cdot 10^{-2}$\\ \hline 
300--350 & $1.66$ 	 & $\pm 0.02$ 	 & $\pm^{0.05}_{0.05}$& $\pm 0.03$ 	  	& $\cdot 10^{-2}$\\ \hline 
350--400 & $7.56$ 	 & $\pm 0.11$ 	 & $\pm^{0.23}_{0.23}$& $\pm 0.14$ 	  	& $\cdot 10^{-3}$\\ \hline 
400--470 & $3.08$ 	 & $\pm 0.05$ 	 & $\pm^{0.10}_{0.10}$& $\pm 0.06$ 	  	& $\cdot 10^{-3}$\\ \hline 
470--550 & $1.16$ 	 & $\pm 0.03$ 	 & $\pm^{0.04}_{0.04}$& $\pm 0.02$ 	  	& $\cdot 10^{-3}$\\ \hline 
550--650 & $3.82$ 	 & $\pm 0.15$ 	 & $\pm^{0.16}_{0.17}$& $\pm 0.07$ 	  	& $\cdot 10^{-4}$\\ \hline 
650--750 & $1.24$ 	 & $\pm 0.09$ 	 & $\pm^{0.05}_{0.06}$& $\pm 0.02$ 	  	& $\cdot 10^{-4}$\\ \hline 
750--900 & $2.96$ 	 & $\pm 0.35$ 	 & $\pm^{0.14}_{0.14}$& $\pm 0.06$ 	  	& $\cdot 10^{-5}$\\ \hline 
\pho900--1100 & $7.23$ 	 & $\pm 1.78$ 	 & $\pm^{0.39}_{0.38}$& $\pm 0.14$ 	  	& $\cdot 10^{-6}$\\ \hline 
\end{tabular}

   }
\caption{
The inclusive prompt photon cross section with systematic and statistical uncertainties for the region $0.6 \leq |\etag| < 1.37$
}
\label{tab:inclusivePhoton_second}
\end{table}

\begin{table}[ht]
\centering
   {\small
   \renewcommand{\arraystretch}{1.3}
\begin{tabular}{|c|ccccc|}
\hline 
$\ETg$ range [\GeV] &  $\mathrm{d}\sigma^{\gamma} / \mathrm{d} \ETg$ & Stat. Unc. & Sys. Unc. & Lumi. Unc. & [pb/\GeV] \\ \hline 
25--35   & $4.42$ 	 & $\pm 0.02$ 	 & $\pm^{0.55}_{0.54}$& $\pm 0.08$ 	  	& $\cdot 10^{2}$\\ \hline 
35--45   & $1.34$ 	 & $\pm 0.01$ 	 & $\pm^{0.11}_{0.11}$& $\pm 0.03$ 	  	& $\cdot 10^{2}$\\ \hline 
45--55   & $4.82$ 	 & $\pm 0.05$ 	 & $\pm^{0.29}_{0.28}$& $\pm 0.09$ 	  	& $\cdot 10^{1}$\\ \hline 
55--65   & $2.15$ 	 & $\pm 0.02$ 	 & $\pm^{0.10}_{0.10}$& $\pm 0.04$ 	  	& $\cdot 10^{1}$\\ \hline 
65--75   & $1.07$ 	 & $\pm 0.01$ 	 & $\pm^{0.05}_{0.04}$& $\pm 0.02$ 	  	& $\cdot 10^{1}$\\ \hline 
75--85   & $5.77$ 	 & $\pm 0.06$ 	 & $\pm^{0.25}_{0.25}$& $\pm 0.11$ 	  	& \\ \hline 
\pho85--105  & $2.69$ 	 & $\pm 0.02$ 	 & $\pm^{0.13}_{0.13}$& $\pm 0.05$ 	  	& \\ \hline 
105-125 & $1.02$ 	 & $\pm 0.01$ 	 & $\pm^{0.05}_{0.05}$& $\pm 0.02$ 	  	& \\ \hline 
125--150 & $4.38$ 	 & $\pm 0.02$ 	 & $\pm^{0.26}_{0.27}$& $\pm 0.08$ 	  	& $\cdot 10^{-1}$\\ \hline 
150--175 & $1.89$ 	 & $\pm 0.01$ 	 & $\pm^{0.13}_{0.13}$& $\pm 0.04$ 	  	& $\cdot 10^{-1}$\\ \hline 
175--200 & $8.98$ 	 & $\pm 0.10$ 	 & $\pm^{0.69}_{0.69}$& $\pm 0.17$ 	  	& $\cdot 10^{-2}$\\ \hline 
200--250 & $3.48$ 	 & $\pm 0.03$ 	 & $\pm^{0.29}_{0.30}$& $\pm 0.07$ 	  	& $\cdot 10^{-2}$\\ \hline 
250--300 & $1.09$ 	 & $\pm 0.01$ 	 & $\pm^{0.10}_{0.10}$& $\pm 0.02$ 	  	& $\cdot 10^{-2}$\\ \hline 
300--350 & $3.76$ 	 & $\pm 0.08$ 	 & $\pm^{0.40}_{0.41}$& $\pm 0.07$ 	  	& $\cdot 10^{-3}$\\ \hline 
350--400 & $1.52$ 	 & $\pm 0.05$ 	 & $\pm^{0.18}_{0.19}$& $\pm 0.03$ 	  	& $\cdot 10^{-3}$\\ \hline 
400--470 & $5.11$ 	 & $\pm 0.22$ 	 & $\pm^{0.63}_{0.69}$& $\pm 0.10$ 	  	& $\cdot 10^{-4}$\\ \hline 
470--550 & $1.27$ 	 & $\pm 0.10$ 	 & $\pm^{0.18}_{0.20}$& $\pm 0.02$ 	  	& $\cdot 10^{-4}$\\ \hline 
550--650 & $2.71$ 	 & $\pm 0.40$ 	 & $\pm^{0.47}_{0.50}$& $\pm 0.05$ 	  	& $\cdot 10^{-5}$\\ \hline 
\end{tabular}

   }
\caption{
The inclusive prompt photon cross section with systematic and statistical uncertainties for the region $1.56 \leq |\etag| < 1.81$.
}
\label{tab:inclusivePhoton_third}
\end{table}

\begin{table}[ht]
\centering
   {\small
   \renewcommand{\arraystretch}{1.3}
\begin{tabular}{|c|ccccc|}
\hline 
$\ETg$ range [\GeV] &  $\mathrm{d}\sigma^{\gamma} / \mathrm{d} \ETg$ & Stat. Unc. & Sys. Unc. &Lumi. Unc. & [pb/\GeV] \\ \hline 
25--35   & $9.47$ 	 & $\pm 0.02$ 	 & $\pm^{1.06}_{1.03}$& $\pm 0.18$ 	  	& $\cdot 10^{2}$\\ \hline 
35--45   & $2.84$ 	 & $\pm 0.01$ 	 & $\pm^{0.21}_{0.21}$& $\pm 0.05$ 	  	& $\cdot 10^{2}$\\ \hline 
45--55   & $1.04$ 	 & $\pm 0.01$ 	 & $\pm^{0.06}_{0.06}$& $\pm 0.02$ 	  	& $\cdot 10^{2}$\\ \hline 
55--65   & $4.48$ 	 & $\pm 0.02$ 	 & $\pm^{0.18}_{0.18}$& $\pm 0.09$ 	  	& $\cdot 10^{1}$\\ \hline 
65--75   & $2.16$ 	 & $\pm 0.01$ 	 & $\pm^{0.07}_{0.07}$& $\pm 0.04$ 	  	& $\cdot 10^{1}$\\ \hline 
75--85   & $1.18$ 	 & $\pm 0.01$ 	 & $\pm^{0.03}_{0.03}$& $\pm 0.02$ 	  	& $\cdot 10^{1}$\\ \hline 
\pho85--105  & $5.37$ 	 & $\pm 0.03$ 	 & $\pm^{0.13}_{0.13}$& $\pm 0.10$ 	  	& \\ \hline 
105--125 & $2.05$ 	 & $\pm 0.01$ 	 & $\pm^{0.04}_{0.04}$& $\pm 0.04$ 	  	& \\ \hline 
125--150 & $8.29$ 	 & $\pm 0.03$ 	 & $\pm^{0.17}_{0.17}$& $\pm 0.16$ 	  	& $\cdot 10^{-1}$\\ \hline 
150--175 & $3.32$ 	 & $\pm 0.07$ 	 & $\pm^{0.07}_{0.07}$& $\pm 0.06$ 	  	& $\cdot 10^{-1}$\\ \hline 
175--200 & $1.52$ 	 & $\pm 0.01$ 	 & $\pm^{0.04}_{0.04}$& $\pm 0.03$ 	  	& $\cdot 10^{-1}$\\ \hline 
200--250 & $5.41$ 	 & $\pm 0.03$ 	 & $\pm^{0.15}_{0.15}$& $\pm 0.10$ 	  	& $\cdot 10^{-2}$\\ \hline 
250--300 & $1.42$ 	 & $\pm 0.02$ 	 & $\pm^{0.05}_{0.05}$& $\pm 0.03$ 	  	& $\cdot 10^{-2}$\\ \hline 
300--350 & $4.18$ 	 & $\pm 0.09$ 	 & $\pm^{0.17}_{0.18}$& $\pm 0.08$ 	  	& $\cdot 10^{-3}$\\ \hline 
350--400 & $1.35$ 	 & $\pm 0.05$ 	 & $\pm^{0.06}_{0.07}$& $\pm 0.03$ 	  	& $\cdot 10^{-3}$\\ \hline 
400--470 & $3.87$ 	 & $\pm 0.19$ 	 & $\pm^{0.20}_{0.21}$& $\pm 0.07$ 	  	& $\cdot 10^{-4}$\\ \hline 
470--550 & $7.17$ 	 & $\pm 0.76$ 	 & $\pm^{0.38}_{0.40}$& $\pm 0.14$ 	  	& $\cdot 10^{-5}$\\ \hline 
550--650 & $1.08$ 	 & $\pm 0.25$ 	 & $\pm^{0.07}_{0.07}$& $\pm 0.02$ 	  	& $\cdot 10^{-5}$\\ \hline 
\end{tabular}

   }
\caption{
The inclusive prompt photon cross section with systematic and statistical uncertainties for the region $1.82 \leq |\etag| < 2.37$.
}
\label{tab:inclusivePhoton_fourth}
\end{table}

\FloatBarrier

\printbibliography

\newpage
\begin{flushleft}
{\Large The ATLAS Collaboration}

\bigskip

G.~Aad$^\textrm{\scriptsize 87}$,
B.~Abbott$^\textrm{\scriptsize 114}$,
J.~Abdallah$^\textrm{\scriptsize 65}$,
O.~Abdinov$^\textrm{\scriptsize 12}$,
B.~Abeloos$^\textrm{\scriptsize 118}$,
R.~Aben$^\textrm{\scriptsize 108}$,
M.~Abolins$^\textrm{\scriptsize 92}$,
O.S.~AbouZeid$^\textrm{\scriptsize 138}$,
N.L.~Abraham$^\textrm{\scriptsize 150}$,
H.~Abramowicz$^\textrm{\scriptsize 154}$
% ...
\bigskip
\\
% ...
$^{12}$ Institute of Physics, Azerbaijan Academy of Sciences, Baku, Azerbaijan\\
% ...
$^{65}$ University of Iowa, Iowa City IA, United States of America\\
% ...
$^{87}$ CPPM, Aix-Marseille Universit{\'e} and CNRS/IN2P3, Marseille, France\\
% ...
$^{92}$ Department of Physics and Astronomy, Michigan State University, East Lansing MI, United States of America\\
% ...
$^{108}$ Nikhef National Institute for Subatomic Physics and University of Amsterdam, Amsterdam, Netherlands\\
% ...
$^{114}$ Homer L. Dodge Department of Physics and Astronomy, University of Oklahoma, Norman OK, United States of America\\
% ...
$^{118}$ LAL, Univ. Paris-Sud, CNRS/IN2P3, Universit{\'e} Paris-Saclay, Orsay, France\\
% ...
$^{138}$ Santa Cruz Institute for Particle Physics, University of California Santa Cruz, Santa Cruz CA, United States of America\\
% ...
$^{150}$ Department of Physics and Astronomy, University of Sussex, Brighton, United Kingdom\\
% ...
$^{154}$ Raymond and Beverly Sackler School of Physics and Astronomy, Tel Aviv University, Tel Aviv, Israel\\
% ...
\end{flushleft}






\end{document}